# Impact of *h*-index on authors ranking: A comparative analysis of Scopus and WoS

Parul Khurana[1] and Kiran Sharma[2]

[1] *parul11183@gmail.com*
School of Computer Applications, Lovely Professional University, Phagwara, 144401, Punjab (India)

[2] *kiran.sharma@northwestern.edu*
Department of Chemical & Biological Engineering, Northwestern University, Evanston, 60208, Illinois (USA)

**Abstract**
In academia, the research performance of the faculty members is either evaluated by the number of publications or the number of citations. Most of the time *h*-index is widely used during the hiring process or the faculty performance evaluation. The calculation of the *h*-index is shown in various databases; however, there is no recent or systematic evidence about the differences between them. In this study, we compare the difference in the *h*-index compiled with Scopus and Web of Science (WoS) with the aim of analyzing the ranking of the authors within a university. We analyze the publication records of 350 authors from Monash University (Australia). We also investigate the discipline wise variation in the authors ranking. 31% of the author's profiles show no variation in the two datasets whereas 55% of the author's profiles show a higher count in Scopus and 9% in WoS. The maximum difference in *h*-index count among Scopus and WoS is 3. On average 12.4% of publications per author are unique in Scopus and 4.1% in WoS. 53.5% of publications are common in both Scopus and WoS. Despite larger unique publications in Scopus, there is no difference shown in the Spearman correlation coefficient between WoS and Scopus citation counts and *h*-index.

**Introduction**
Citations analysis plays a crucial role while evaluating the research performance of an individual in the academic community, that is why it acts as a key tool in scientometrics (Cronin, Snyder & Atkins, 1997; Bornmann, 2017). Along with the citations, a number of publications and *h*-index also have a strong stand in research evaluation. In order to perform the fair evaluation of an individual within a university/institution, funding bodies, scientific society, etc., it is the essential requirement that the consider scientometrics parameters should be field, discipline, and time normalized (Waltman, 2016). However, with the rapid increase in the number of scholarly databases or libraries like Google Scholar, Scopus, Web of Science, Dimension, PubMed, etc., the choice of database consideration has become tedious (Bakkalbasi et al., 2006; Falagas et al., 2008; Meho & Yang, 2007; Mongeon & Paul-Hus, 2016; Martín-Martín, 2018).

In bibliometric, *h*-index which is based on the number of publications and citations received on those publications, considered as one of the important measures used to evaluate the individual work quality, impact, influence, and importance (Bar-Ilan, Levene & Lin, 2007; Bar-Ilan, 2008). Researchers have shown the use and importance of *h*-index measures while calculating the ranking of authors, universities, the impact of a journal, etc. (Costas & Bordons, 2007; Bornmann & Daniel, 2009; Torres-Salinas, Lopez-Cózar & Jiménez-Contreras, 2009; Vieira & Gomes, 2009). Dunaiski et al. have evaluated the bias and performance of the authors over a range of citation; however, no significant differences between the globalized and averaged variants based on citations was found (Dunaiski, Geldenhuys & Visser, 2019). Different approaches have been used in literature to analyze the author's ranking. Authors have also used the page rank algorithm on the author co-citations network to rank the authors (Ding et al., 2009; Nykl, Campr & Ježek, 2015; Dunaiski, Visser & Geldenhuys, 2016; Dunaiski, Geldenhuys & Visser, 2018). Usman et al. have shown in their research the analysis of various assessment parameters like *h*-index, citations, publications, authors per paper, *g*-index, *hg*-index, *R*-index, e-index, etc. to evaluate the authors ranking (Usman, Mustafa & Afzal, 2020). The aim of the study is to highlight the impact of the *h*-index on the author's profile while ranking/evaluating the authors' performance within a university. So, we have examined the research contributions, in terms of the number of publications, citations, and *h*-index of authors of Monash University from Scopus and WoS. This study aims to answer the following research questions:

1) What is the impact of *h*-index on authors ranking based on different databases?
2) How ranking of authors of Monash University varies with the number of publication and citations provided by WoS and Scopus across disciplines?
3) How much is the deviation among disciplines when we take the difference of *h*-index provided by WoS and Scopus?

**Data description and filtration**

*Data selection*

Data selection has two important aspects, *first*, what research question we are going to answer, and *second*, what is the required approach to answering the question. The goal is to study the ranking of authors based on the number of publications, citations and *h*-index computed from Scopus and WoS. The very first challenge was the selection of the authors. On what basis an author should be selected was the major concern. To get detailed information about the scholarly data of any author we need authentic information. Indexing databases like Scopus or WoS tracks the author identity information with Author ID or Researcher ID or Orcid ID. There were two ways to approach this: one was to look for any open-source dataset like Kaggle, etc. and the other was to look manually at the university websites. Finally, we found that Monash University, a public research university in Australia has provided the profiles of 6316 persons associated with the university at different designations and in different subject areas/disciplines. There were three associated benefits with this dataset:

- First, all the profiles were sorted on the basis of the last name of persons.
- Second, authors Orcid ID, Researcher ID, and Scopus ID was mentioned.
- Third, a search tab was given on the website to filter the profiles with at least 5 years' or 10 years' work with the university.

Then, we started searching the profiles of the persons manually by opening all the profiles one by one. A further requirement of the research question was to identify those profiles which have all of three ID's: Orcid ID, Researcher ID, and Author ID. Orcid ID is a digital identifier and is used to uniquely identify authors across different platforms. Researcher ID is a digital identifier used by WoS to maintain the database of authors. Author ID is used by Scopus for the unique identification of authors. After checking each and every profile manually on the website of Monash University, we have considered the profile carrying all three IDs (see flowchart in Figure 1 (left panel)). To critically evaluate the identified research question, we have recorded the subject area/discipline of persons along with all three IDs (Orcid, Researcher, and Author). In the end, a sample of 350 persons from various subject areas/disciplines was finalized and is used to analyze the identified research question.

*Data filtration*

In order to perform the analysis, we have visited the profiles of 350 authors listed on the Monash University website (https://research.monash.edu/en/persons/), which is freely available. Extracting different IDs (Orcid, Researcher, and Author) to avoid author ambiguity at any stage, later on, was the initial step of scrapping the data. So, the information extracted from Monash University regarding 350 authors was (see Figure 1 (left panel)):
- Author name
- Authors Orcid, Researcher, and Scopus ID
- Authors disciplines or subject categories

Further, we extracted the following information from both Scopus and WoS on May 2020 using respective API's (see Figure 1 (right panel):
- Author name, affiliation, *h*-index
- Detailed records of the number of publications and citations received on those publications for all 350 authors
- *doi's* of all publications

A total of 24988 documents was downloaded for all authors from Scopus and 24095 from WoS. To maintain the uniqueness among downloaded data, we considered all the records with *doi* numbers only. Thus, we filtered the number of documents in Scopus with *doi* as 23073 (92.3%) and 20064 (83.6%) in WoS. Further, we have filtered the *common (*both in Scopus and WoS) and *unique* (either Scopus or WoS*)* documents across both indexing databases. We have found that Scopus has 4353 (17.8%) unique documents, and WoS has 1419 (5.8%) unique documents. 18720 (76.4%) of documents were common in both indexing databases. Hence for further analysis and statistics, we have considered *common + unique* documents across Scopus and WoS as a final contribution of an author.

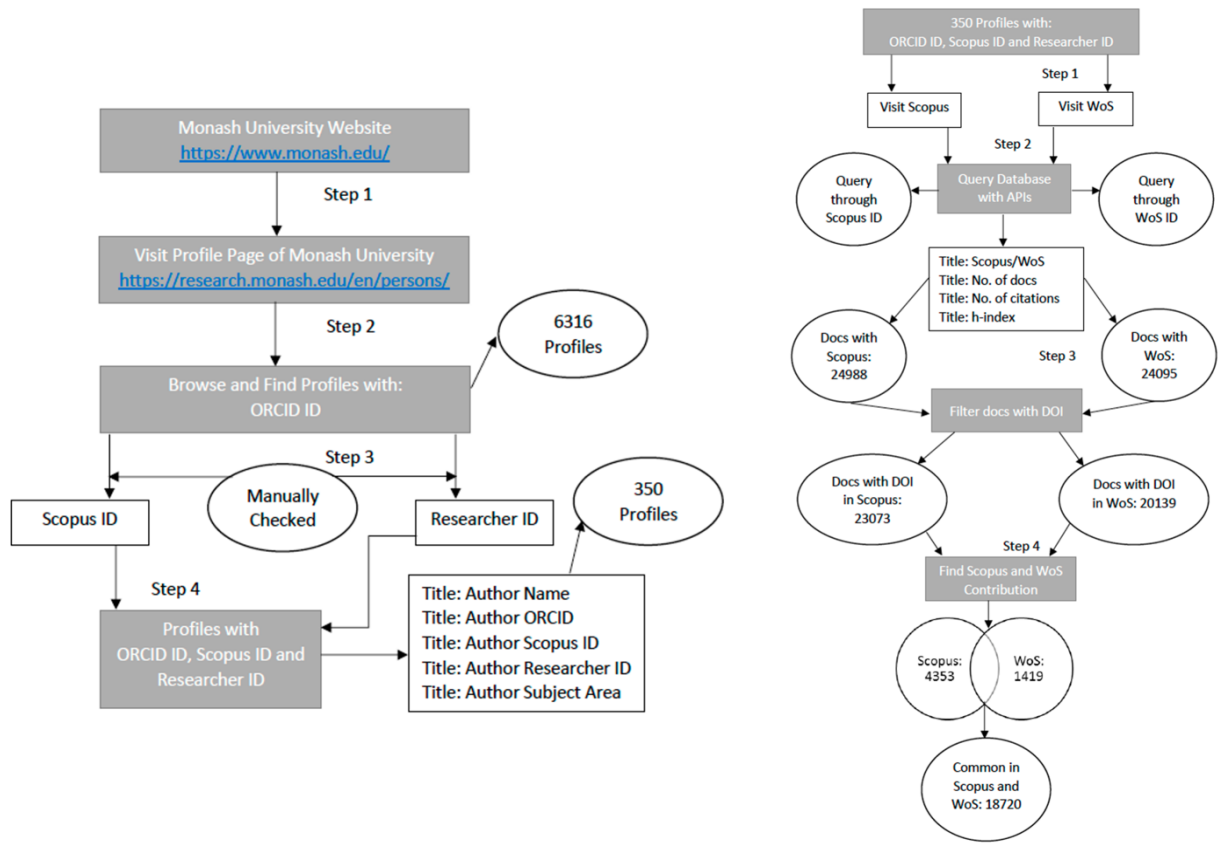

**Figure 1. Flowchart describing the steps for (left panel) visiting the author's profiles and (right panel) data extraction and filtration from Scopus and WoS.**

## Results

*Authors ranking based on h-index*

Figure 2 shows the ranking of 350 authors based on the *h*-index computed in Scopus and WoS. Scopus *h*-index is arranged in descending order and the corresponding WoS *h*-index is plotted. Ranks of WoS *h*-index is showing large fluctuation at a lower level in comparison to Scopus. 31% of authors have zero variation in *h*-index in both databases whereas 45% shares the difference of one, 19% of two, and 5% of three values of *h*-index. Table 1 shows the number of authors, publications, citations, and average *h*-index of 350 authors aggregated in eight disciplines of Monash University in Scopus and WoS. Figure 3 shows the variation in the standard deviation of the author's *h*-index aggregated over disciplines. The standard deviation is performed on the difference of Scopus and WoS *h*-index. *Agriculture and Environment, Biochemistry and Molecular Biology, and Physics, Chemistry and Mathematics* disciplines show the large amount of deviation in authors *h*-index. Table 2 displays the discipline wise count of the number of authors based on the *h*-index. *Biochemistry and Molecular Biology, Engineering, and Health and Medical Sciences* have authors with *h*-index>50.

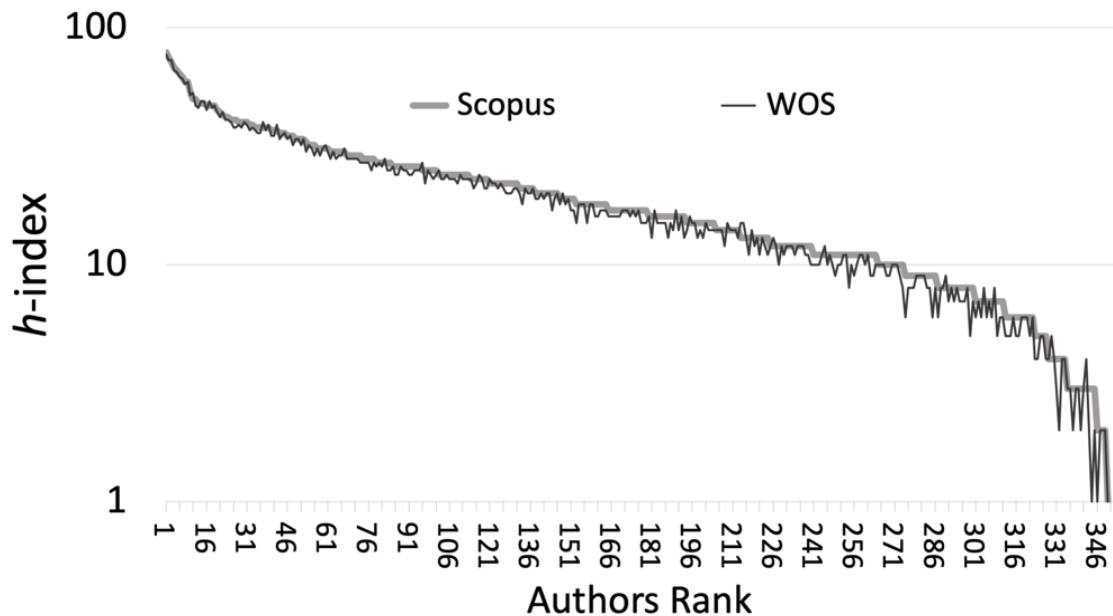

Figure 2. The plot shows the author's ranking based on the *h*-index of 350 authors in Scopus and WoS. Authors with less than 10 *h*-index show a lot of variations and the high value of the *h*-index shows less variation in both the databases.

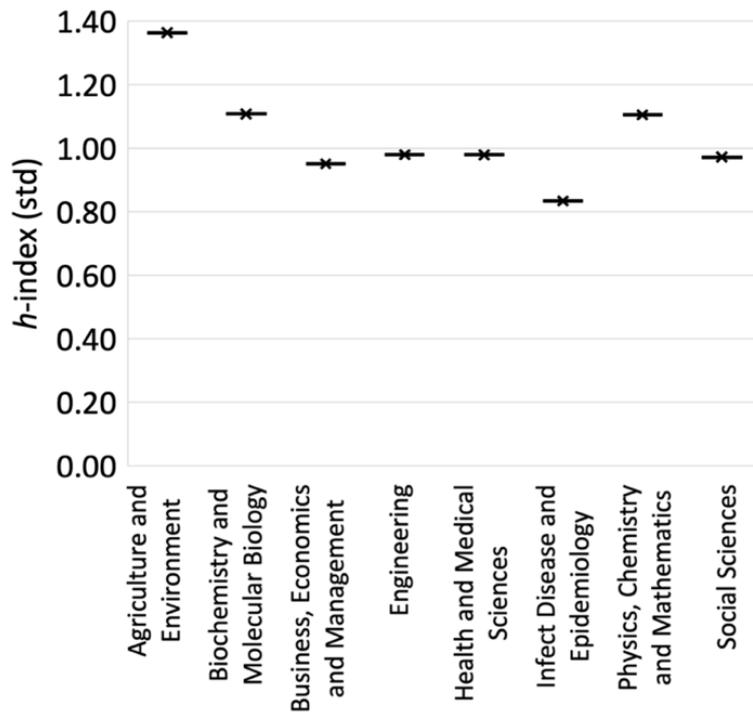

**Figure 3. Discipline wise deviation in the authors *h*-index between Scopus and WoS.**

**Table 1. The number of publications, citations, and average *h*-index of 350 authors aggregated in eight disciplines of Monash University in Scopus and WoS.**

| Disciplines | # Authors | Scopus | | | WoS | | |
|---|---|---|---|---|---|---|---|
| | | Total Pub | Total Citations | Avg h-index | Total Pub | Total Citations | Avg h-index |
| Agriculture and Environment | 18 | 1068 | 35866 | 20.2 | 1004 | 32687 | 19.4 |
| Biochemistry and Molecular Biology | 83 | 6439 | 271758 | 24.7 | 6373 | 262767 | 24.2 |
| Business, Economics and Management | 7 | 236 | 11899 | 11.6 | 175 | 9268 | 10.9 |
| Engineering | 56 | 4937 | 119572 | 19.6 | 4576 | 109132 | 18.7 |
| Health and Medical Sciences | 88 | 6529 | 240966 | 20.0 | 6477 | 224655 | 19.3 |
| Infect Disease and Epidemiology | 8 | 401 | 11038 | 15.9 | 398 | 10070 | 15.0 |
| Physics, Chemistry and Mathematics | 28 | 2198 | 65514 | 21.8 | 2051 | 64202 | 21.3 |
| Social Sciences | 62 | 3180 | 65420 | 14.5 | 3041 | 58756 | 13.3 |

**Table 2. Discipline wise count of the number of authors based on *h*-index provided by Scopus (S) and WoS (W).**

| *Disciplines* | *No. of Authors* | | | | | | | | | | | |
|---|---|---|---|---|---|---|---|---|---|---|---|---|
| | *h ≤10* | | *11≥ h ≤20* | | *21≥ h ≤30* | | *31≥ h ≤40* | | *41≥ h ≤50* | | *h >50* | |
| | S | W | S | W | S | W | S | W | S | W | S | W |
| Agriculture and Environment | 4 | 5 | 7 | 6 | 4 | 4 | 1 | 1 | 2 | 2 | 0 | 0 |
| Biochemistry and Molecular Biology | 9 | 13 | 29 | 27 | 25 | 24 | 11 | 11 | 4 | 3 | 5 | 5 |
| Business, Economics and Management | 3 | 3 | 3 | 3 | 1 | 1 | 0 | 0 | 0 | 0 | 0 | 0 |
| Engineering | 14 | 18 | 21 | 19 | 11 | 10 | 5 | 5 | 3 | 2 | 2 | 2 |
| Health and Medical Sciences | 26 | 28 | 29 | 28 | 16 | 16 | 9 | 8 | 5 | 4 | 3 | 4 |
| Infect Disease and Epidemiology | 1 | 2 | 5 | 5 | 1 | 0 | 1 | 1 | 0 | 0 | 0 | 0 |
| Physics, Chemistry and Mathematics | 5 | 6 | 10 | 9 | 7 | 7 | 4 | 4 | 2 | 2 | 0 | 0 |
| Social Sciences | 24 | 29 | 23 | 19 | 11 | 10 | 3 | 4 | 1 | 0 | 0 | 0 |

*Discipline wise comparative analysis of authors rank based on number of publications and citations*

Figure 4 shows the discipline wise distribution of the number of publications (common and unique) in Scopus and WoS. A large variation in the number of unique publications is mainly analyzed in *Business, Economics and Management (27%), Engineering (18%), Physics, Chemistry and Mathematics (13%), and Social Sciences (13%).* The variations in the number of unique publications in *Biochemistry and Molecular Biology* is 13.5% in Scopus and 3.6% in WoS. On average Scopus has a large number of unique publications for all disciplines. The variation in the number of publications with *doi* is larger in *Engineering* (23%) in WoS and in *Infect Disease and Epidemiology* (13.8%) in Scopus. Similarly, the proportion of common publication is higher in Scopus (15%) for *Physics, Chemistry and Mathematics* and in WoS (19%) for *Biochemistry and Molecular Biology*. Table 3 shows the description of the number of publications (total, with *doi*, unique and common) for 350 authors aggregated in eight disciplines.

Figure 5 shows the variation in the ranking of the authors across disciplines based on the number of publications and citations. For all disciplines, Spearman rank was calculated for all associated authors and correlation was calculated between the authors rank in Scopus and authors rank in WoS. Similarly, the rank between Scopus and WoS citations was calculated. The analysis shows that the author's rank does not get much affected by the number of citations; however, there is variation in terms of the number of publications in some disciplines. The Spearman rank correlation coefficient of *Business, Economics and Management, Engineering, Health and Medical Sciences, and Social Sciences* varies for the number of publications.

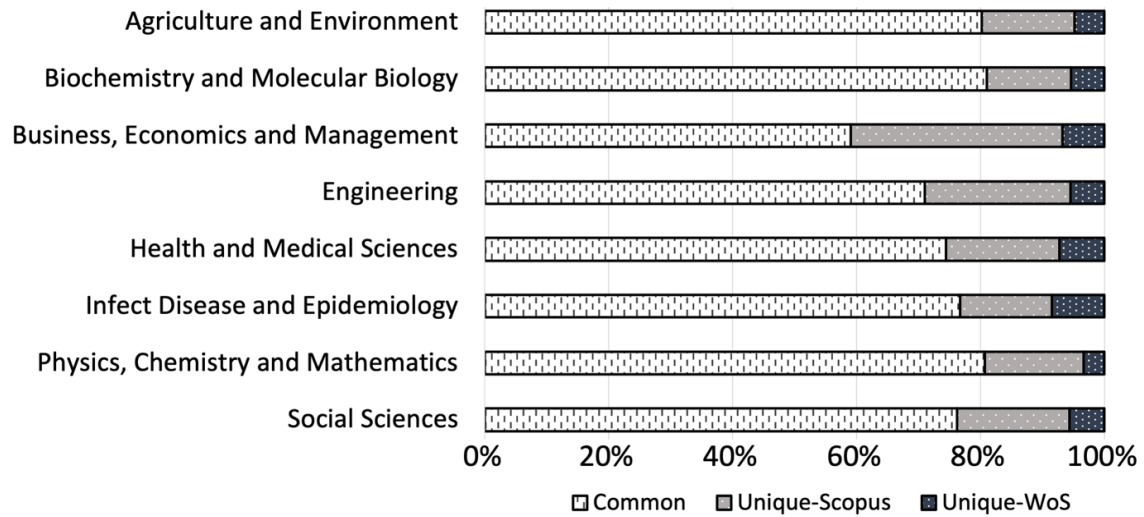

**Figure 4. Discipline wise distribution of the number of publications (common and unique) in Scopus and WoS. Unique publications are those which only appear either in Scopus or in WoS.**

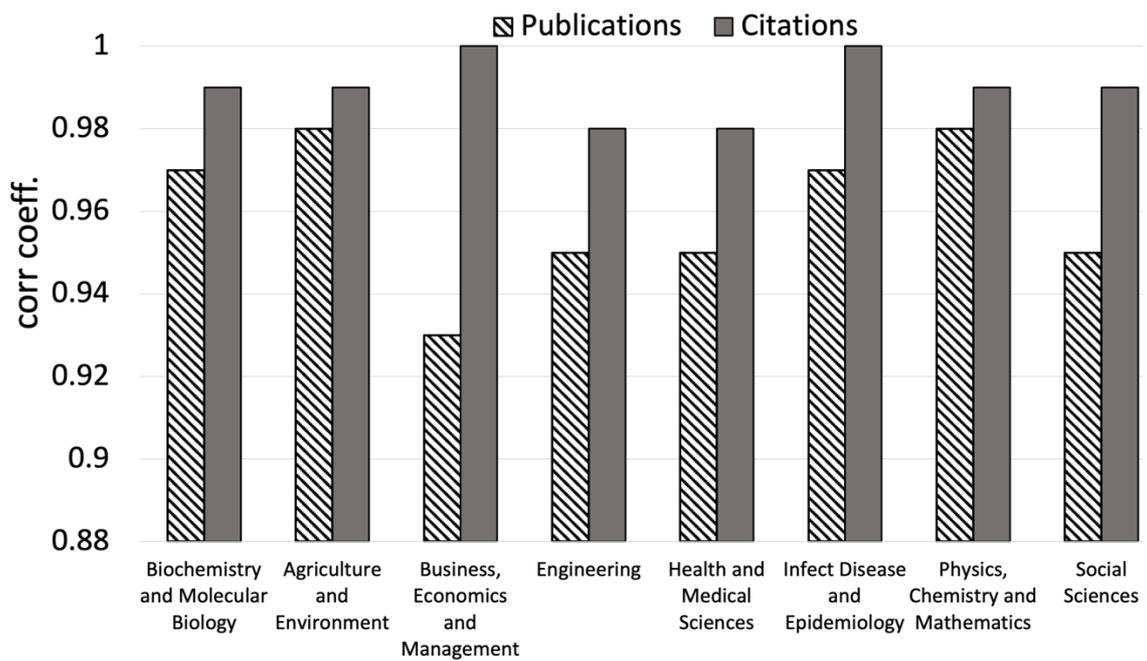

**Figure 5. shows the discipline wise variation in Spearman rank correlation coefficients of authors rank based on the number of publications and citations provided by Scopus and WoS.**

**Table 3. Detailed calculation of the number of publications (total, with *doi*, unique and common) for 350 authors aggregated in eight disciplines of Monash University in Scopus and WoS.**

| Disciplines | Scopus | | | WoS | | | Total Common Pub |
|---|---|---|---|---|---|---|---|
| | Total Pub | Total Pub With doi | Total Unique Pub | Total Pub | Total Doc With doi | Total Unique Pub | |
| Agriculture and Environment | 1068 | 989 | 156 | 1004 | 883 | 50 | 833 |
| Biochemistry and Molecular Biology | 6439 | 6084 | 872 | 6373 | 5558 | 346 | 5212 |
| Business, Economics and Management | 236 | 221 | 81 | 175 | 156 | 16 | 140 |
| Engineering | 4937 | 4360 | 1086 | 4576 | 3525 | 251 | 3274 |
| Health and Medical Sciences | 6529 | 6016 | 1187 | 6477 | 5303 | 474 | 4829 |
| Infect Disease and Epidemiology | 401 | 346 | 56 | 398 | 322 | 32 | 290 |
| Physics, Chemistry and Mathematics | 2198 | 2088 | 344 | 2051 | 1817 | 73 | 1744 |
| Social Sciences | 3180 | 2969 | 571 | 3041 | 2575 | 177 | 2398 |

**Conclusion**

This study provides insight and meaningful implications regarding research/performance evaluation of faculties within a university-based on the number of publications, citations, and *h*-index. Organizations usually use the scholarly data and analysis based on those data to evaluate the performance of individual; however, due to the availability of multiple sources like Google Scholar, Web of Science, Scopus, Dimension, Microsoft Academic Graph, etc., it made the job bit tedious. It is always the question of debate that which scholarly data one should consider while performing the evaluation. In this study, we used Scopus and WoS data to perform the analysis on faculty research of Monash University. The study analyses the small sample of 350 authors (publications, citations, and *h*-index) of Monash University across eight disciplines. The size and importance of the discipline is also a major factor and it varies across the universities. Despite this limitation, the study contributes significantly to research, as we are first time showing the impact of the *h*-index on authors ranking in different databases across eight disciplines and showed empirically how the use of different databases provides a more comprehensive picture of an author's research rank.

The results also showed the variation in authors ranking based on the number of publications; however, the ranking does not get affected by the number of citations. So, it is difficult to choose the appropriate evaluation parameter. Either of the choice can create a dilemma for promotion committees and science policymakers. So, we analyzed the ranking based on the *h*-index; however, the variation in *h*-index across different databases also made the decision cumbersome. Bar-Ilan (2008) mentioned in his study that the *h*-index only provides partial information. We recommend further explore the capabilities and limitations of *h*-index in terms of the author's discipline, area of research, and date of joining of the institution.